# Interlayer Conductance of Graphene with Multiple Transfer Process

Yin Sun, Xintong Zhang, Longyan Wang, Lining Zhang[a], Salahuddin Raju and Mansun Chan

*Department of ECE, Hong Kong University of Science and Technology, Clear Water Bay, Kowloon, Hong Kong*

Electrical properties of multi-layer graphene are subject to variations due to random interlayer alignments. In this work we reported graphene interlayer conductance without special layer aligning. Ohmic contacts between two graphene layers are observed with resistance variations of more than one order. With Raman spectroscopy we identify that the lattice angle between twisted graphene layers is the key variation source. The angular dependence and temperature dependence of the interlayer conductance suggest that a phonon assistant tunneling mechanism is valid for the interlayer transport of graphene prepared by multiple transfer process. We finally derive that the multi-layer graphene resistance shows an exponential-like distribution due to the random interlayer misalignments.

Vertically stacked van der Waals structures[1-4] have drawn growing interests for their potentials in variety of assembly devices. Within this category multi-layer graphene weights in a wide range of applications[5-6], e.g. transparent electrode for display and lighting, functional electrode of transistors, metallic interconnect with higher current-carrying capacity and so on. Novel devices are also obtained by inserting a functional layer between two graphene layers[7-8]. Starting from the chemical vapor deposition (CVD) of monolayer graphene, so far the layer-by-layer transfer method (multiple transfer process) is still the most widely used process to form the graphene stacks[9-10] due to the capability for large scale application[11]. As can be expected, many stacking patterns besides of the Bernal one are possible[12] without special controls over alignments. There will be different amounts of sliding and twisting between the honeycomb lattices of two adjacent layers. The random layer misalignment with the multilayer graphene process is a potential cause of property variations. Some theoretical studies[13] suggest that the graphene interlayer conductance strongly depends on the twisted angle and may show differences of more than 10 orders of magnitude with the angles spanning from 0 to 30 (angles of 30 to 60 are equivalent due to the symmetry). Under this scenario, the applications of the layer-by-layer transferred graphene will be hindered due to intolerant variations. Later it is theoretically revisited that this angle dependence may be significantly relaxed after considering impacts of phonon scattering in the interlayer transport[14]. Based on that, the variations of multilayer graphene will be significantly reduced. Though some experimental results[15] of high quality exfoliated graphene layers are available in literatures to support the later theory, variations of the most technology relevant multi-layer graphene prepared by the multiple transfer process are yet unknown. As a possibly dominant variation source, the interlayer conductance of CVD graphene needs to be figured out, however, is still unavailable from the literatures.

In this letter we reported the electrical conductance of CVD graphene interlayers, their variations and impacts on the multi-layer graphene resistance. Graphene contacts formed in different two-layer graphene stacks without any intentional alignments, and those formed in one stack but within different graphene grains are characterized to reveal the conductance variations. We confirm the angle dependence generally agrees with predictions of the phonon assisted tunneling mechanism. Finally we discuss the variations in the multi-layer graphene interconnect.

---

[a] **Electronic mail:** eelnzhang@ust.hk

Electrical contacts to individual graphene layer are needed to characterize the interlayer conductance. We have designed a test structure for this purpose, as shown in Fig.1 (a). A dual layer device is prepared on silicon substrate with 100nm thermally grown silicon dioxide. The graphene is synthesized by chemical vapor deposition (CVD) on copper and then integrated into this device via the widely used wet transfer process. After transferring the first graphene layer, we define the metal contacts of Ti (10nm)/Au (50nm) by the thermal evaporation and lift-off process. Graphene is patterned using oxygen plasma. To avoid etching of the first layer when patterning the upper layer, an etching mask layer with $Al_2O_3$ (20nm) is formed with atomic layer deposition. With a square graphene-graphene contact window defined by another photolithography and hard mask etching by FHD-5, a second graphene layer is transferred on top and patterned with $O_2$ plasma. High selectivity of FHD-5 on $Al_2O_3$ over $SiO_2$ mitigates the surface roughness of the device substrate. Finally another electrode formation step is done to contact the second graphene layer. Multiple graphene-graphene contacts with different areas are designed and grouped closely, and several groups are formed within one transfer process. A few samples are prepared successively with random twisting angles between the two graphene layers. Fig.1 (b) shows the Raman spectroscopy of a sample graphene stack in comparisons with that of the monolayer. Increased intensity of the G peak and 2D peak indicates the success of stack formations. By comparing the 2D peak characteristics in Raman spectroscopy of the one-layer and two-layer graphene[16], we obtain the twisting angles between two graphene layers.

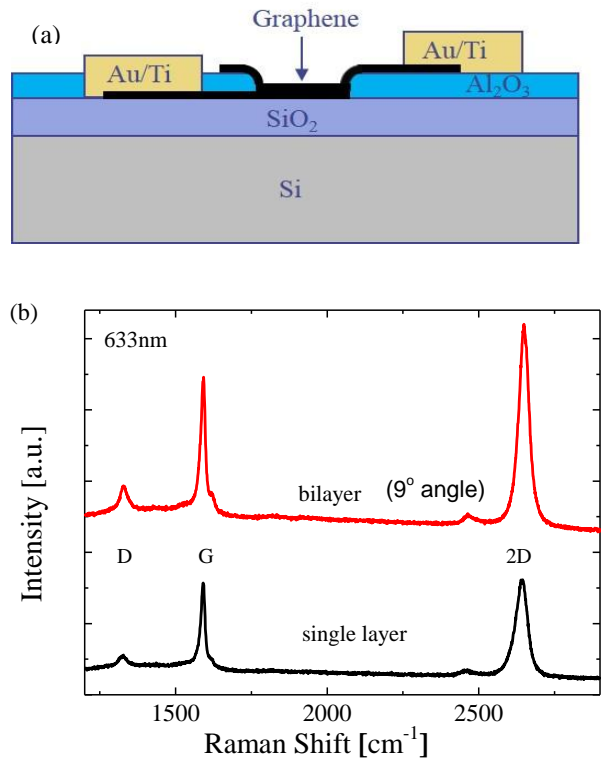

Figure 1: (a) Schematic of the test structures for graphene interlayer conductance; (b) Raman spectroscopy of single-layer graphene and double-layer graphene. A twisted angle of 9 degree between two graphene layers is extracted.

We characterize all the devices in atmosphere using Agilent 4156C for their current-voltage properties. Hall measurements of the transferred graphene show p-type dopings and the Fermi energies ($E_F$) are all around 4.8 eV. Although there is no electron momentum perpendicular to the graphene plane, scatterings induced electron transfer and their electric field dependence promise a resistor-like interlayer conduction. As both the graphene intralayer resistance and the metal-graphene contact resistances do not show voltage dependence, a linear current-voltage characteristic is expected with resistance of $R_{total}$. Fig.2 (a) plots the ohmic property of test structures of different area within one group. The interlayer resistance is obtainable by subtracting the other two resistances. For that, we include the transmission line measurement (TLM) structures having a column of two-point probe patterns with varying distances, from which we calculate the graphene sheet resistance $\rho_{sh}$ and metal-graphene contact resistivity $\rho_c$ at room temperature. After accounting for number of squares $N$ and contact width $W$, the interlayer resistance $R_{bi}$ is thus found by:

$$R_{bi} = R_{total} - \rho_{sh}N - 2\frac{\rho_c}{W} \qquad (1)$$

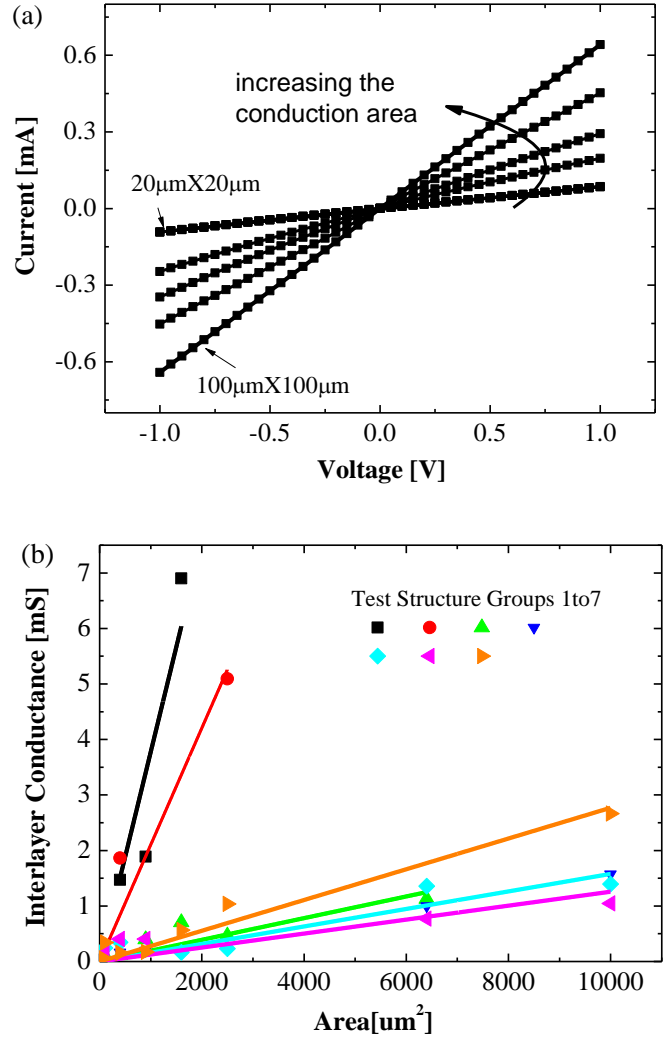

Figure 2: (a) I-V characteristics of test structures of different area (20x20 µm$^2$, 30x30 µm$^2$, 40x40 µm$^2$, 80x80 µm$^2$, 100x100 µm$^2$) within one group. The inset shows the height histogram of the one-layer to the two-layer regions. The two peaks are fitted by the sum (solid line) of two Gaussian functions (dashed lines). (b) The extracted graphene interlayer conductance versus graphene contact area. Results from seven groups of test structures are included.

The above procedure is repeated for all the samples. The average graphene interlayer distance is around 0.5 nm measured with atomic force microscopy (AFM) at the layer transition regions of the test structure. It is larger than the equilibrium distance of 0.34nm in the Bernal stacked graphite. Fig.2 (b) summaries the graphene interlayer conductance in different groups of test structures. The lines are linear fittings to experimental data to guide eyes. Within one group, the contact conductance is almost linearly proportional to the contact area with certain amount of variations. With these results, we exclude the possibility of significant interference on the interlayer transport from the metal residuals since they are mostly non-uniform and random distributed. At the same time, the graphene edge effects[17] are not observed in our transferred graphene stack. As a result, the interlayer current conduction can be assumed to be almost uniformly distributed over the entire contact area. Among different groups, the conductance per area varies more than one order of magnitude. One highly possible reason is the random twisting angels between the graphene grain domains in the stack. The CVD graphene is not single crystalline and different graphene grains are rotated and meet at the grain boundary[18]. It is also possible that one grain in the top graphene layer covers two or more grains in the bottom graphene layer and vice versa. For simplicity we assume one group of graphene stacks has one unique twisting angle.

After characterizing the test structures over a large amount of samples, we find that the total variations are up to around 50 times. It is interesting that the order of variations agree quite well to that predicted by a phonon assisted interlayer tunneling theory[14] in the range of 0 to 30 degree angles. Briefly, the involvement of flexural phonons[19] of the bilayer graphene weakens the confinements of momentum and energy conservations in the interlayer transport. When the twisting angles approach those under which significant reductions of pi-orbitals overlapping happen, like 15 degree, the flexural phonons scattering weight over the umklapp process and reduce the angle dependence of the interlayer tunneling. The interlayer conductivity (S/μm$^2$) is given by:

$$G_{bi} = 2\pi e^2 g^2 n E_F A_0 / M_c \omega \pi^2 \hbar^4 v_F^4 \qquad (2)$$

where $M_c$ is the carbon mass, $A_0$ is the area of the two atom cell, $v_F$ is the Fermi velocity, $g$ describes the strength of electron-phonon interactions which is interlayer distance dependent, $n$ is the number of flexural phonons with frequency of $\omega$, and wave vectors of these phonons twist the electron momentum during the interlayer transition. With the phonon numbers and frequency well known, the only fitting parameter $g$ is determined from one data point of conductance versus angle. We use a device with 18 degree angle to determine the fitting parameter and get the theoretical angular dependence of the conductance, as shown in Fig.3 (a). Another device with 9 degree is right sitting on top of the model curve which verifies that the interlayer transport of our CVD transferred graphene are described reasonably well with the phonon assistant tunneling theory. The measured data of interlayer conductance is described well by the model as shown with the scattered data in Fig.3 (a). The conductance with small twisting angles is about 4-5 orders lower than that of the graphite[20], and is around 1 order lower than that of exfoliated graphene[15], which is possibly due to the large graphene interlayer distance. Any PMMA residuals from the wet transfer process[21] may hinder the relaxation of graphene layers to the ideal equilibrium stage. The error bars indicate co-existence of other second order variation sources like the metal residuals.

Number of phonons increases with temperatures and accordingly the interlayer conductance is expected to have a positive temperature dependence. We perform characterizations of the graphene stack structures and TLM structures, and the same procedure is carried out to extract the interlayer conductance under different temperatures. Fig.3 (b) plots the interlayer resistance where the expected trend is clearly observed. At the same time, the temperature dependence predicted by the phonon assistant tunneling model Eq. (2) is plotted as the curve in Fig.3 (b). Close agreements with the experimental results serve another evidence for the phonon assistant transport mechanism.

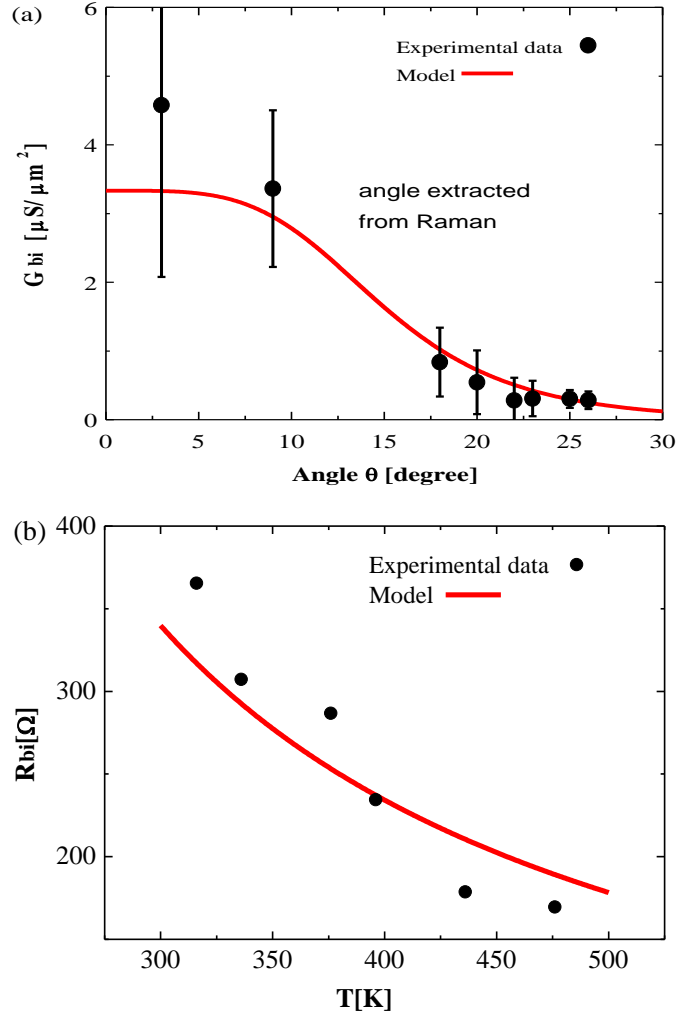

Figure 3: (a) Dependence of interlayer graphene conductance on the twisted angle θ. Experimental results (symbols) are well described by a phonon assistant tunneling theory (line). Error bars are included for test structures within one group. (b) Temperature dependence of the interlayer resistance of one test structure. Experimental results (symbols) are in agreements with the theory prediction (line). .

Validations of the phonon assisted tunneling model for the layer-by-layer transferred CVD graphene have important implications, e.g. on graphene interconnect. Assuming a segment of top-contact interconnect formed with multiple transfer graphene and a random twisting angle between different layers, we can obtain the total interconnect resistance with a resistor network model[22]. We first consider two extreme cases, i.e. one case with all layers aligned and another case with two adjacent layers twisted by 30°, and plot the interconnect resistance versus the number of layers in Fig.4. The interlayer conductance is Fig.3 (a) is used as an example, and the interconnect length and width is assumed to be 80μm and 10μm. In the worst case, increasing the number of layers is not effective to reduce the total interconnect resistance. Monte Carlo simulations are performed to emulate this randomness. We find an exponential-type resistance distribution of the 4-layer graphene interconnect in the inset of Fig.4. Both the higher mean resistance and the variations are caused by the random interlayer misalignments. However, the probability to get a resistance close to that of the aligned system is also high due to the relative higher conductivity of double layer graphene with small angles as seen in Fig.3 (a). Techniques such as those for interlayer distance reduction and graphene doping, are expected to further improve the multi-layer interconnect.

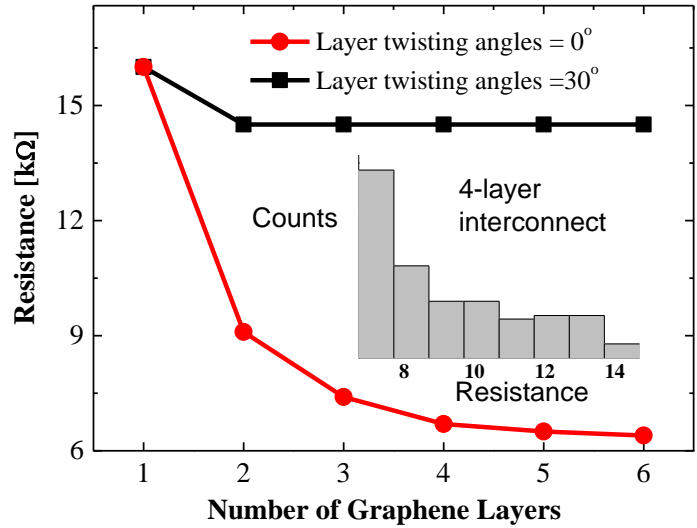

Figure 4: Resistance of a multilayer graphene interconnect (top contact) is reduced by larger number of layers, however, with variations from the random twisted angles. The inset shows an exponential-type distribution in the range of minimum to maximum conductivity.

In conclusion, we found that the interlayer conductance of transferred CVD graphene is well described by the phonon assistant tunneling mechanism. The random twisted angle between graphene layers is mainly responsible for the variations in interlayer conductance. It acts as one key variation source in devices with multi-layer graphene structures. The results should have implications on various applications where multi-layer graphene is implemented, e.g. transparent graphene electrode for transistors.

The work was supported by the Area of Excellence Grant under the Contract No. AOE/P-04/08 from the University Grant Council of Hong Kong.